\documentclass{article}

\usepackage{arxiv}

\usepackage[utf8]{inputenc} 
\usepackage[T1]{fontenc}    
\usepackage{hyperref}       
\usepackage{url}            
\usepackage{booktabs}       
\usepackage{amsfonts}       
\usepackage{nicefrac}       
\usepackage{microtype}      
\usepackage{lipsum}
\usepackage{graphicx}
\usepackage{float}

\title{Electrochemical double layer capacitor using a microemulsion electrolyte}

\author{
  Fraser R Hughson\thanks{Corresponding Author} \\
  School of Chemical and Physical Sciences\\
  Victoria University of Wellington\\
  Wellington 6140, New Zealand\\
  \texttt{fraser.hughson@vuw.ac.nz}\\
   \And
 Rohan Borah \\
  School of Mathematical and Chemical Sciences
\\
  The University of Newcastle\\
 Newcastle, NSW 2308, Australia \\
  \texttt{rohan.borah@newcastle.edu.au} \\
  \AND
 Thomas Nann \\
  School of Mathematical and Chemical Sciences
\\
  The University of Newcastle\\
 Newcastle, NSW 2308, Australia \\ 
  \texttt{thomas.nann@newcastle.edu.au} \\
}

\begin{document}
\maketitle

\begin{abstract}
The use of aqueous electrolytes in energy storage devices is traditionally limited by the voltage stability window of water at 1.23 V. We present the use of a microemulsion based electrolyte which, although mostly water by mass, has a voltage stability window of up to 5 V. This allows the cost and safety benefits of aqueous electrolytes to finally be realised in high-voltage systems. Supercapacitors constructed using this electrolyte were able to achieve and maintain a capacitance of ~40 F/g and an energy density ~40 Wh/kg of  with a Coulombic efficiency of 99\%  for over 10,000 cycles on activated carbon.  
\end{abstract}


\section{Introduction}

In the diverse energy storage landscape, supercapacitors offer the advantage of fast charge and discharge making them ideal for high power applications such as accelerating vehicles or voltage stabilisation \cite{gonzalez_review_2016}. The current standard electrolyte used in supercapacitors consists of an organic solvent (such as acetonitrile or propylene carbonate) and a quaternary ammonium salt. This combination has good electrochemical stability but is costly as well as potentially hazardous \cite{zhao_review_2015}. Using a water-based electrolyte is an attractive option, however, above a potential of 1.23V it is thermodynamically favourable for water to split into oxygen and hydrogen gasses. This means that under most circumstances aqueous supercapacitors are limited to around 1V at maximum. Traditional supercapacitors achieve voltages of between 2.7-3V, above which it is the electrode material (activated carbon) that begins to degrade rather than the electrolyte \cite{yang_carbon_2019}. Here we present the first use of a microemulsion electrolyte, with water as the major component, in a supercapacitor with which we can achieve 2.7V. By breaking the 1.23V barrier while remaining cheap and sustainable, we anticipate this electrolyte to have significant impact on aqueous electrochemical energy storage technologies and their eventual commercialisation.

Many attempts have been made to extend the electrochemical window of water so that it may be used in high energy density applications such as mobile phone batteries. Recently water-in-salt electrolytes have become popular \cite{suo_water--salt_2015, leonard_water--salt_2018-1, suo_water--salt_2017-1, rlukatskaya_concentrated_2018} but other strategies also exist such as choosing materials with high over-potentials for the water splitting reactions \cite{shao_design_2018, suarez-guevara_hybrid_2014} or altering the mass balance of the two electrodes \cite{yu_new_2018, feng_effect_2018, cericola_effect_2011}. Microemulsions have been used to perform electrochemistry in the past \cite{peng_electron_2020, rusling_electrochemical_1997-1, rusling_green_2001-1, gao_electrochemical_1998}, however they have never been pursued in supercapacitors. Microemulsions are thermodynamically stable mixtures of two immiscible solvents, usually water and an ‘oil’, where oil is used to refer to any sufficiently hydrophobic liquid. A surfactant and sometimes a cosurfactant are included to allow for the formation of the microemulsion. 

\section{Results and Discussion}
\label{sec:headings}

\begin{figure}
\includegraphics[width=\textwidth]{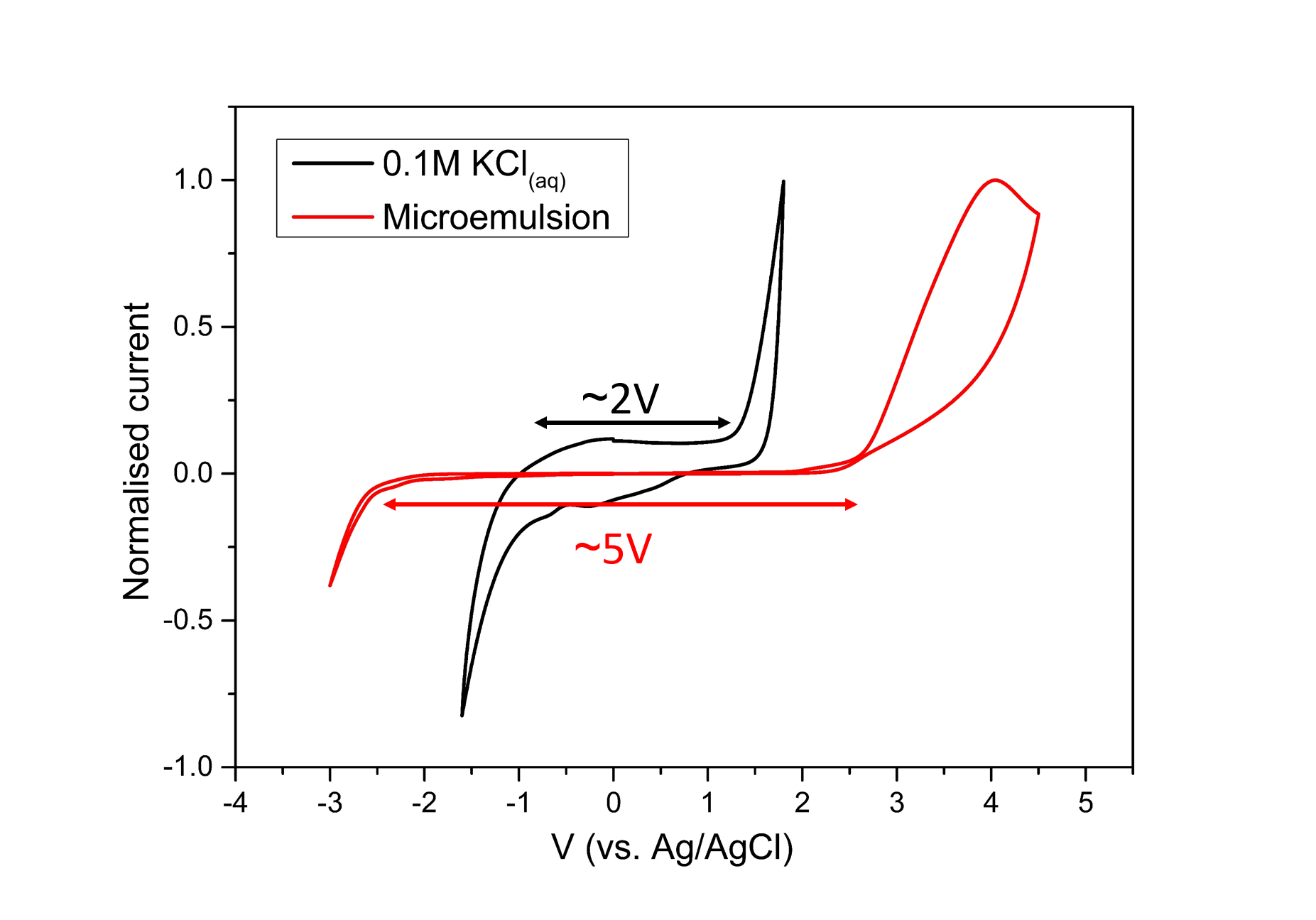}
\centering
  \caption{Cyclic voltammograms of an aqueous electrolyte and a microemulsions electrolyte at 100 mV/s on glassy carbon using an Ag/AgCl reference electrode. The approximate electrochemical windows are indicated. The currents have been normalised to allow for ease of comparison.}
  \label{fig:ME Window}
\end{figure}

Figure \ref{fig:ME Window} shows a cyclic voltammogram of an aqueous solution of 0.1M KCl and a microemulsion electrolyte with 84 wt\%  distilled water, 4 wt\% Sodium dodecyl sulfate (SDS), 9 wt\% n-butanol and 3 wt\% cyclohexane. This was recorded at 100 mV/s on a glassy carbon electrode surface.

It was observed that the electrochemical window of the microemulsion is approximately 5 V compared to the ~2 V of the KCl solution (Figure \ref{fig:ME Window}). The KCl solution showed a window wider than 1.23 V because the water splitting reactions are not favoured on a glassy carbon surface, which causes a high overpotential. This is consistent with other reports of supercapacitors using neutral aqueous solutions being able to achieve 1.6 V \cite{demarconnay_symmetric_2010, xia_symmetric_2012}. 

Following this discovery, different solutions were studied on different working electrode surfaces, in order to establish the role of the components of the microemulsion. Solutions of 0.1M aqueous KCl, a surfactant free microemulsion consisting of 26 wt\% dichloromethane, 36.5 wt\% ethanol and 37.5 wt\% water with 0.1M KCl, a solution of 4.9 wt\% SDS and the SDS based microemulsion described previously were all tested on a glassy carbon electrode (as a hydrophobic surface) and on a platinum electrode (as a hydrophilic surface).

On the platinum surface (Figure SI 1), the electrochemical windows of all solvents are similar with only slight variation between them. This indicates that in order to observe the extended electrochemical window, a hydrophobic surface must be used.

However, on the glassy carbon surface (Figure SI 2), there is a distinct difference between the solutions. Both the surfactant free microemulsion and the 0.1M KCl solution have similar onset potentials for both oxidation and reduction. The solution with just SDS shows a slightly extended electrochemical window compared to just water which agrees with previous results \cite{hou_surfactant_2017}. However, the microemulsion which has both the surfactant and the oil phase is able shows a significantly larger electrochemical window on both the oxidative and reductive scans. 

\begin{figure}
\includegraphics[width=\textwidth]{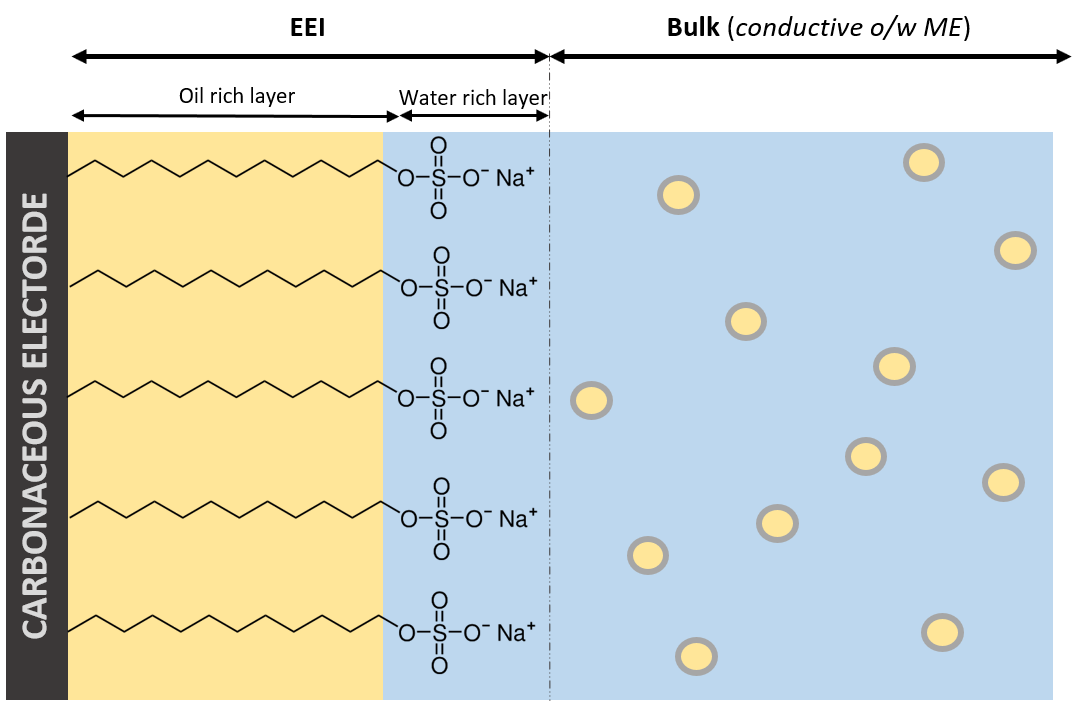}
\centering
  \caption{The proposed mechanism by which the electrochemical window of the electrolyte is extended. As the barrier for diffusion of water through the electrolyte is quite large, no/very little water will reach the electrode surface resulting in no water splitting being observed.}
  \label{fig:EEI}
\end{figure}

Therefore, this large extension of the window can be attributed to the formation of an electrode electrolyte interphase (EEI) on the hydrophobic glassy carbon surface (Figure \ref{fig:EEI}). 

It has been observed previously that the hydrophobic tail of the surfactant arranges itself on the hydrophobic electrode surface \cite{peng_electron_2020}. In this surfactant layer the oil phase is concentrated due to interactions between the non-polar tail of the surfactant and the non-polar oil phase. This is followed by a water rich layer around the polar head group of the surfactant molecules. Due to this oil rich layer, little to no water comes into contact with the electrode surface and so water splitting is effectively suppressed and the voltage window over which the voltage is stable is much wider. 

In order to utilise this extended electrochemical window and to test the ability of the microemulsion electrolyte to function in an energy storage device, it was used in an electrochemical double layer capacitor (EDLC). 

For the supercapacitor electrolyte, sodium chloride (NaCl) was added to the above base microemulsion to achieve a concentration of 0.1 mol/kg. Given that a hydrophobic electrode surface is required activated carbon was chosen as the electrode material over other popular alternatives such as Ruthenium oxides\cite{hu_how_2004} or zeolites \cite{devese_suppressed_2020-1}. All cells were prepared in ambient conditions on a benchtop without degassing of the electrolytes prior to the construction of the device. Cells containing an aqueous NaCl electrolyte (0.1 M) and cells containing an aqueous solution that was 4.9 wt\% SDS were also constructed and tested with galvanostatic charge discharge experiments to test their voltage stability (Figure SI 3).

\begin{figure}
\includegraphics[width=\textwidth]{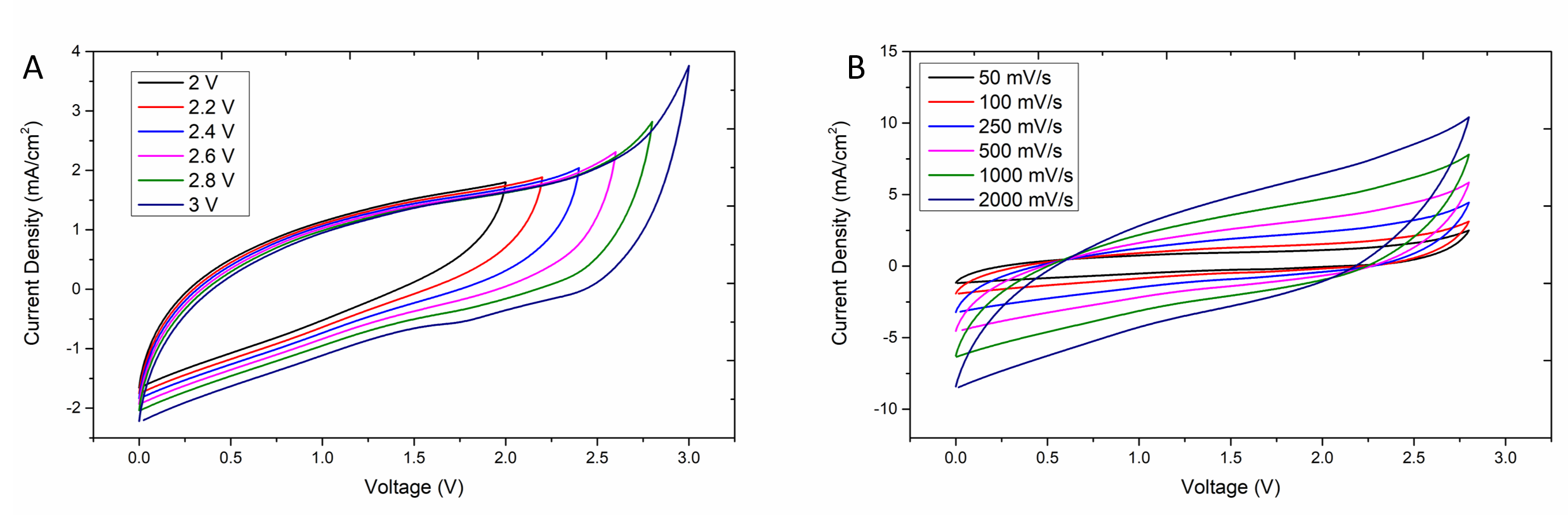}
\centering
  \caption{Cyclic voltammetry of supercapacitors using a 2 electrode set-up. A) Sequential increasing of the upper voltage cut off in 0.2V increments from 2V-3V at 100 mV/s. B) Cyclic voltammograms at various scan speeds using 2.8V as the cut-off. All show deviation from the ideal square wave shape due to the internal resistance of the cell. }
  \label{fig:Cap CVs}
\end{figure}

Cyclic voltammetry was used to investigate the electrochemical window of the cell. Figure \ref{fig:Cap CVs}a shows the cut-off voltage being raised sequentially. From this it can be determined that the maximum voltage of these supercapacitors is approximately 2.7V

Using electrochemical impedance spectroscopy (EIS), the internal resistance of the cell was estimated to be 26 Ohms. Figure SI 3 shows the comparison of a supercapacitor prepared with a purely aqueous electrolyte and a cell prepared with the microemulsion electrolyte. The EIS was repeated on the microemulsion cell after it had completed 10,000 cycles. The aqueous cell had an equal Na\textsuperscript{+} ion concentration as the microemulsion. 

It was observed that the internal resistance of the aqueous cell (approximated by where it touches the x axis on the Nyquist plot in Figure SI 3) is lower than the microemulsion electrolyte. This is consistent with the model presented in Figure 2, where the oil/surfactant layer increases the resistance by reducing the mobility of the charges at the EEI. 

It was also observed that the internal resistance of the cell decreased slightly after 10,000 cycles. This could be attributed to the conditioning processes the cell goes through during its initial cycling improving the wetting of the electrolyte at the electrode. Most importantly, this internal resistance does not increase greatly upon cycling.

\begin{figure}
\includegraphics[width=\textwidth]{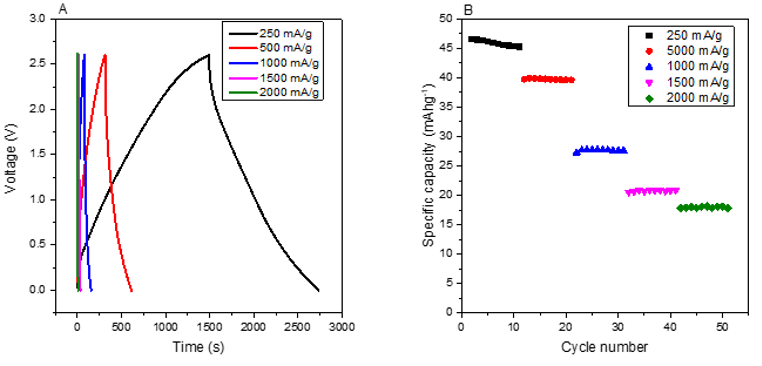}
\centering
  \caption{ A) Galvanostatic cycling of a fresh cell at various rates. B) Capacitance values as a function of the discharge rate. Capacitance values were calculated by It/$\Delta$Vm where I is the current in Amps, t is the time in seconds $\Delta$V is the change in potential (after taking the IR drop into account) and m is the mass of the active material on both electrodes.}
  \label{fig:CDCrates}
\end{figure}

The long term cycling performance of the cells were also investigated. Over 10,000 cycles between 0 and 2.7 V at a rate of 500 mA/g,  very little degradation of the capacitance was observed (Figure 5). Initially the capacitance did decrease, but levelled off around after around 500 cycles and showed roughly constant capacitance from that point forward. An energy density of approximately 40 Wh/kg was achieved during these cycles. The Coulombic efficiency was also high at approximately 99\%. 

To judge the performance of this new electrolyte, cells using a traditional 1M solution of tetrabutyl ammonium tetrafluoroborate (TBAB) in acetonitrile were also constructed using the same activated carbon electrodes. These cells were able to achieve a capacitance of 38-40 F/g when cycled at 500 mA/g. This shows that the the capacitance attained by the acetonitrile cells is similar to that of the microemulsion electrolyte. The rate capability of the acetonitrile electrolyte was better than the microemulsion electrolyte, retaining a high capacitance of 35 F/g whereas the microemulsion electrolyte only achieved 17 F/g. This rate capability could be improved by better matching the surface of the activated carbon to the microemulsion, either by refinement of the pore size or by adjusting the surfactant used and/or component ratios in the composition. 

\begin{figure}
\includegraphics[width=\textwidth]{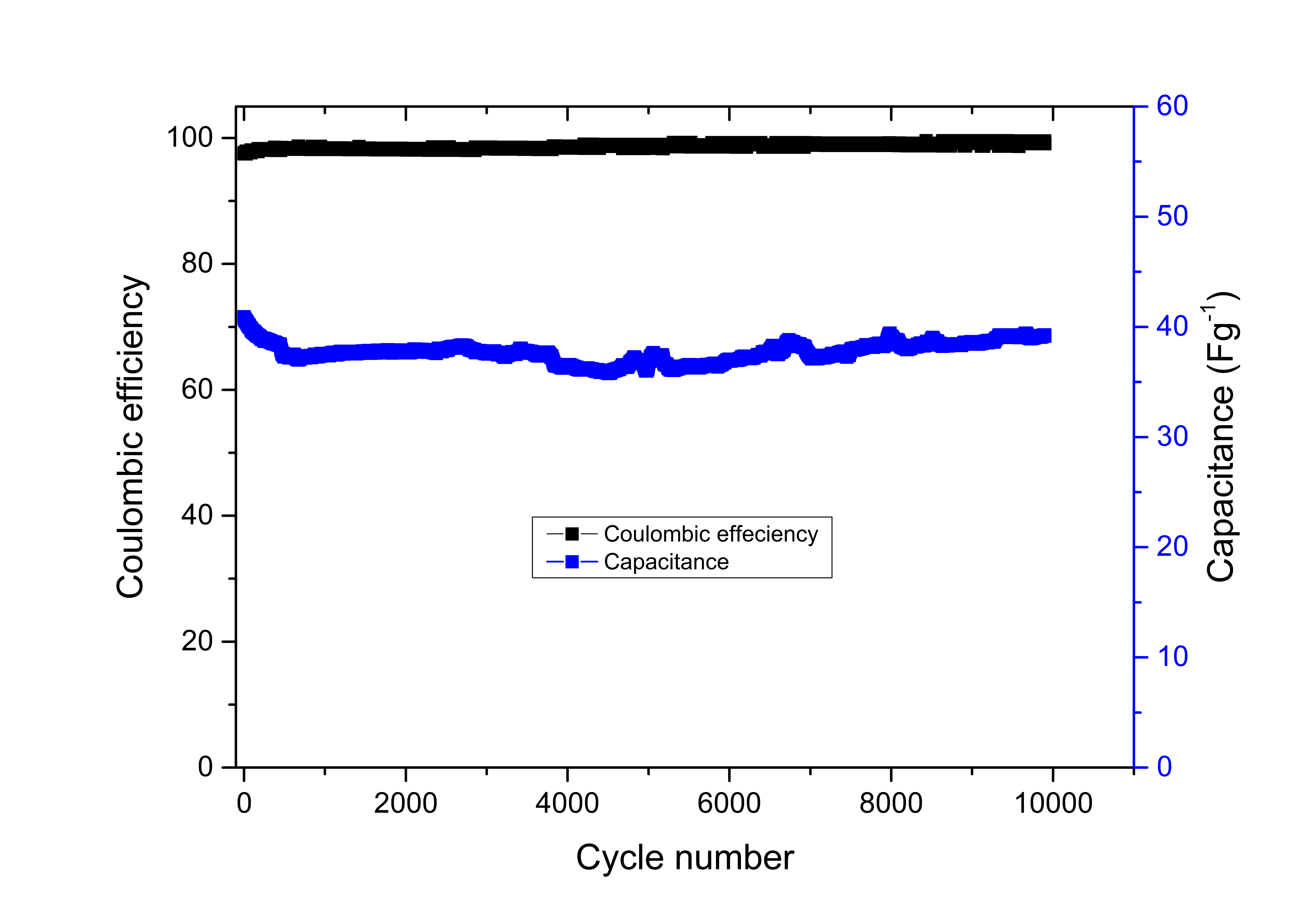}
\centering
  \caption{Coulombic efficiency and capacitance of EDLC over the first 10,000 cycles at a rate of 500 mA/g with upper and lower voltage cut-offs at 2.7 and 0 volts respectively. Coulombic efficiency remained constant at approximately 99\%. After an initial drop, the capacitance value stabilised at approximately 40 F/g and even gradually increased above cycle 500. }
  \label{fig:10kcycles}
\end{figure}

\section{Conclusions}

Using a microemulsion extends the electrochemical window of water to a point where it can be used as a supercapacitor electrolyte with no loss of voltage compared to standard systems based on organic solvents. In this study, we explored the effect of the widened electrochemical window on glassy carbon and platinum electrodes using a microemulsion electrolyte. Furthermore, an electrochemical double-layer capacitor was constructed from activated carbon electrodes and the new electrolyte. A significant increase in voltage was observed despite the electrolyte containing mostly water. Thus, microemulsion electrolytes will enable the fabrication of supercapacitors with similar performance to current technology, using low hazard and low cost materials.

\section{Experimental methods}

\subsection{Materials}

Electrode material slurries were prepared by mixing Activated carbon (85\% by wt. Commodities NZ), 9\% binder (PVDF, MTI Corporation) and 6\% Super‐P conductive carbon (99+\% metals basis, Alfa Aesar). Slurries were ‘doctor-bladed’ onto graphite foil (thickness 0.127 mm, Ceramaterials) and dried in a vacuum oven at 120 Celcius for 12 hours. The specific loading of the active materials was approximately 11 mgcm\textsuperscript{-2}

\subsection{Electrolyte preparation}

The composition of the electrolyte base was kept constant between cells. 84 wt\% distilled water, 4 wt\% Sodium dodecyl sulfate (SDS) (TCI, 99.9\%), 9 wt\% n-butanol (Sigma, reagent grade) and 3 wt\% cyclohexane (Sigma, reagent grade) were mixed in a conical flask until a clear solution was observed indicating the formation of the microemulsion. Occasionally, sonication was utilised to speed up the formation of the microemulsion. After the formation of the microemulsion, NaCl (Fisher scientific) was added to give a concentration of 0.1 mol/kg. The surfactant free microemulsion was prepared in a similar fashion 26 wt\% dichloromethane, 36.5 wt\% ethanol (Sigma, reagent grade) and 37.5 wt\% water with 0.1M KCl (Fisher scientific).

Acetonitrile (Sigma, anhydrous) electrolytes were prepared by mixing acetontirle and tetrabutylammonium tetrafluoroborate (Sigma, 99.9\%) to a concentration of 1M in a nitrogen glovebox.

\subsection{Cell assembly}

Polyether ether ketone (PEEK) cells were used for all electrochemical tests. Custom PEEK rods with glassy carbon rod inserts were used as current collectors. Electrodes were placed inside the cell and glass microfiber (Grade GF/D, Whatman) were used as separators. Approximately 70 microliters of the electrolyte were used to wet the separator. All cell assembly was done on a bench top in ambient conditions.

For acetonitrile cells, assembly was done inside a nitrogen glovebox.

\subsection{Electrochemical testing}

Cyclic voltammetry (CV) experiments on full cells and Pt working electrodes were run using a potentiostat (eDAQ EA 160, ecorder-401), the glassy carbon experiments were done using a biologic VMP3 potentiostat. CVs done to determine electrochemical windows were done in a 3 electrode set up using a platinum-iridium alloy mesh or platinum mesh as the counter electrode, a silver wire as the reference electrode and either glassy carbon or platinum as the working electrode. All solutions were degassed by bubbling nitrogen through them for at least 20 minutes prior to the experiment. CV experiments on full supercapacitor cells were done using the anode as the reference and counter electrodes and the cathode as the working electrode. No degassing was performed prior to electrolytes being loaded into the cells. For acetonitirle cells, cells were taken out of the glovebox and wrapped in parafilm to prevent any exposure to air.

Galvanostatic charge-discharge experiments were performed using a battery analyser system (NEWARE BTS CT-4008-5V10mA-164, MTI Corp.).

\bibliographystyle{unsrt}  
\bibliography{references_Zotero}  

\newpage

\section*{Supporting Information}

\begin{center}
    \textbf{\Huge Electrochemical double layer capacitors using a microemulsion electrolyte}\vspace{0.5ex}
    
    \textit{Fraser R Hughson, Rohan Borah, Thomas Nann*}
\end{center}

\begin{figure}[H]
\includegraphics[width=\textwidth]{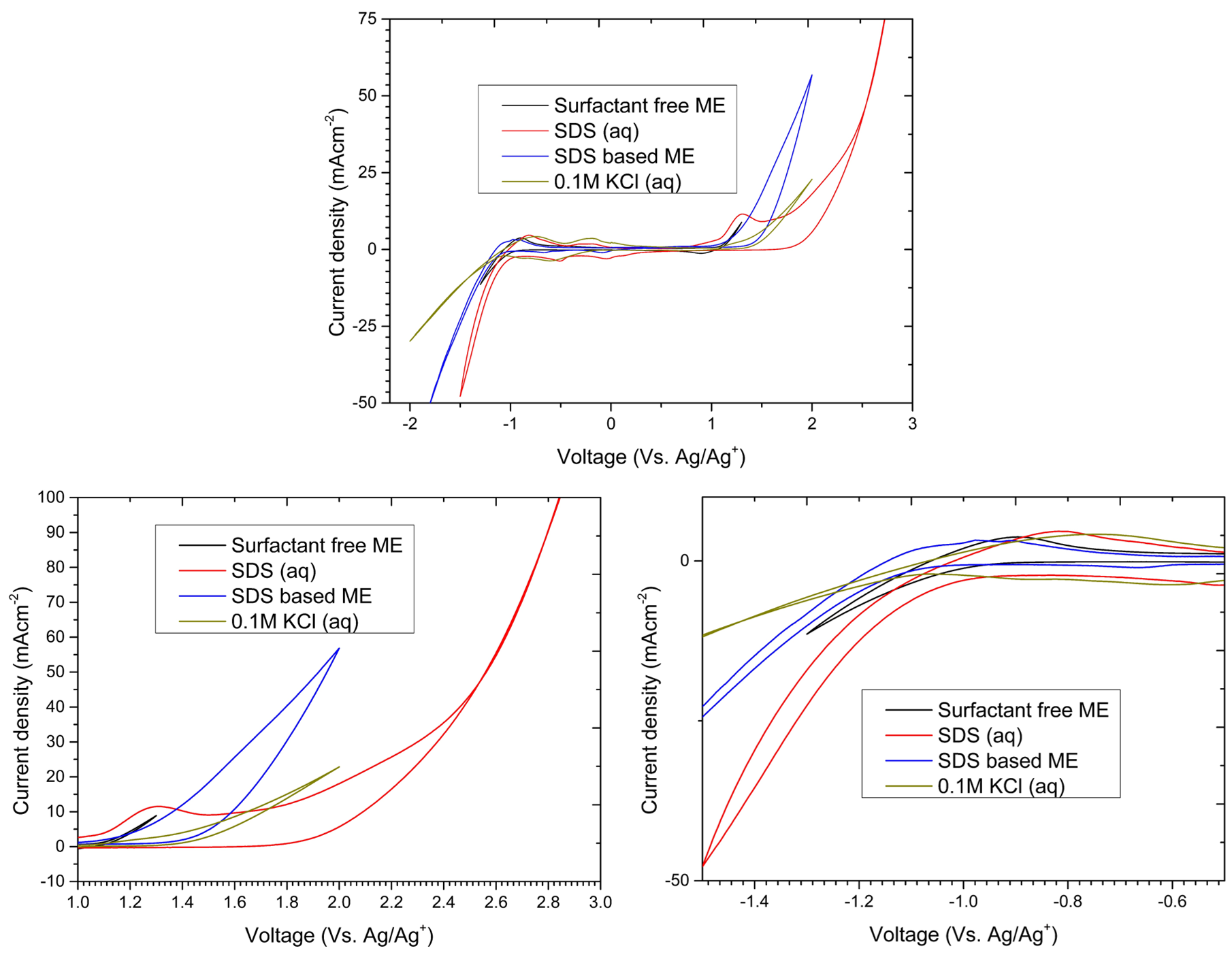}
\centering
  \caption{Windows of the 4 electrolytes: Surfactant free microemulsion, aqueous KCl, aqueous SDS and the SDS based ME on a platinum working electrode. All scans were done at 100 mV/s, each scan shown is the third scan to eliminate any non-equilibrium effects. Each solution was degassed  by bubbling nitrogen through them for at least 20 minutes prior to the experiment  }
  \label{fig:Pt Windows}
\end{figure}

\begin{figure}
\includegraphics[width=\textwidth]{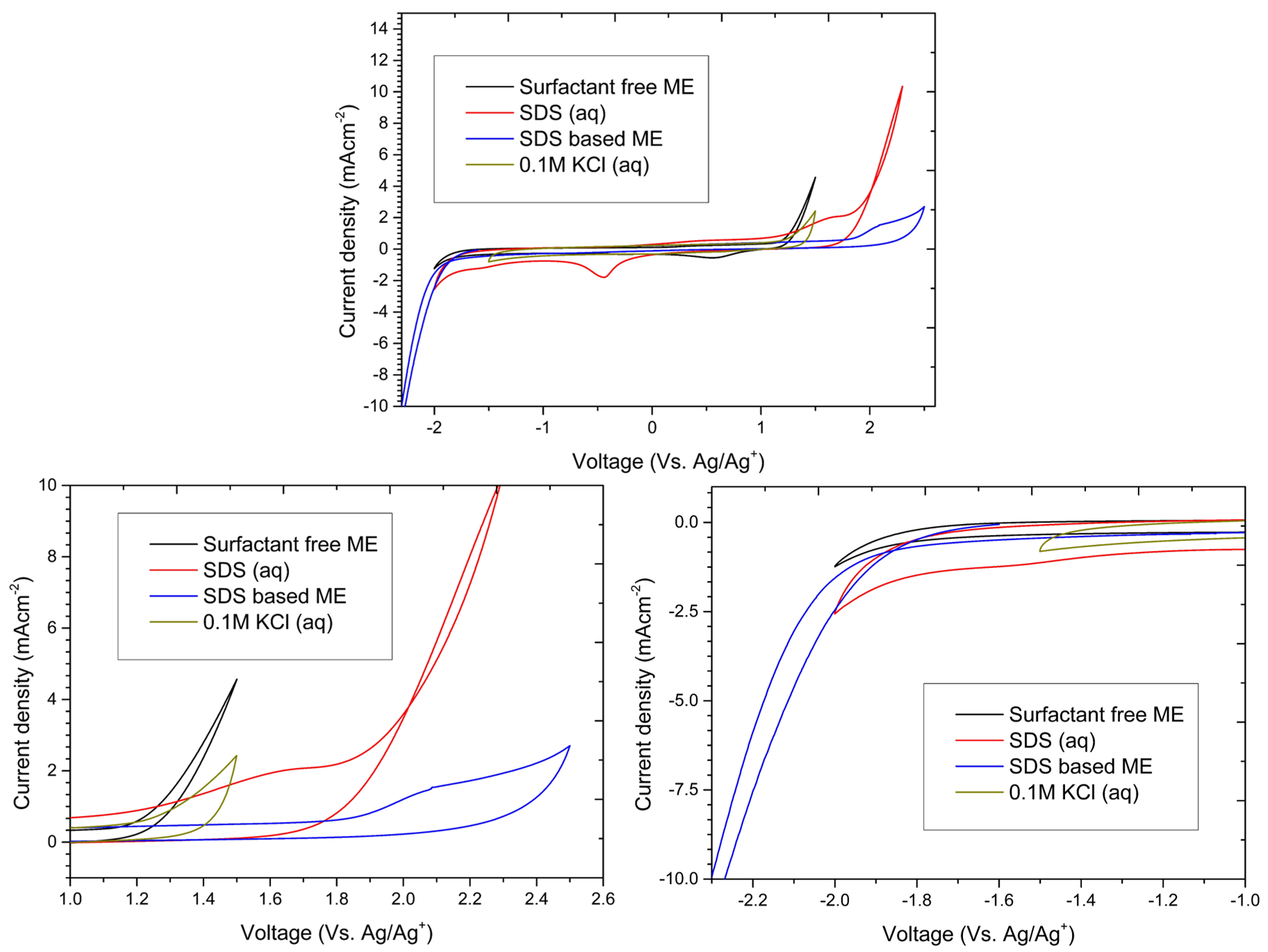}
\centering
  \caption{Windows of the 4 electrolytes: Surfactant free microemulsion, aqueous KCl, aqueous SDS and the SDS based ME on a platinum working electrode. All scans were done at 100 mV/s, each scan shown is the third scan to eliminate any non-equilibrium effects. Each solution was degassed  by bubbling nitrogen through them for at least 20 minutes prior to the experiment }
  \label{fig:Pt Windows}
\end{figure}

\begin{figure}
\includegraphics[width=\textwidth]{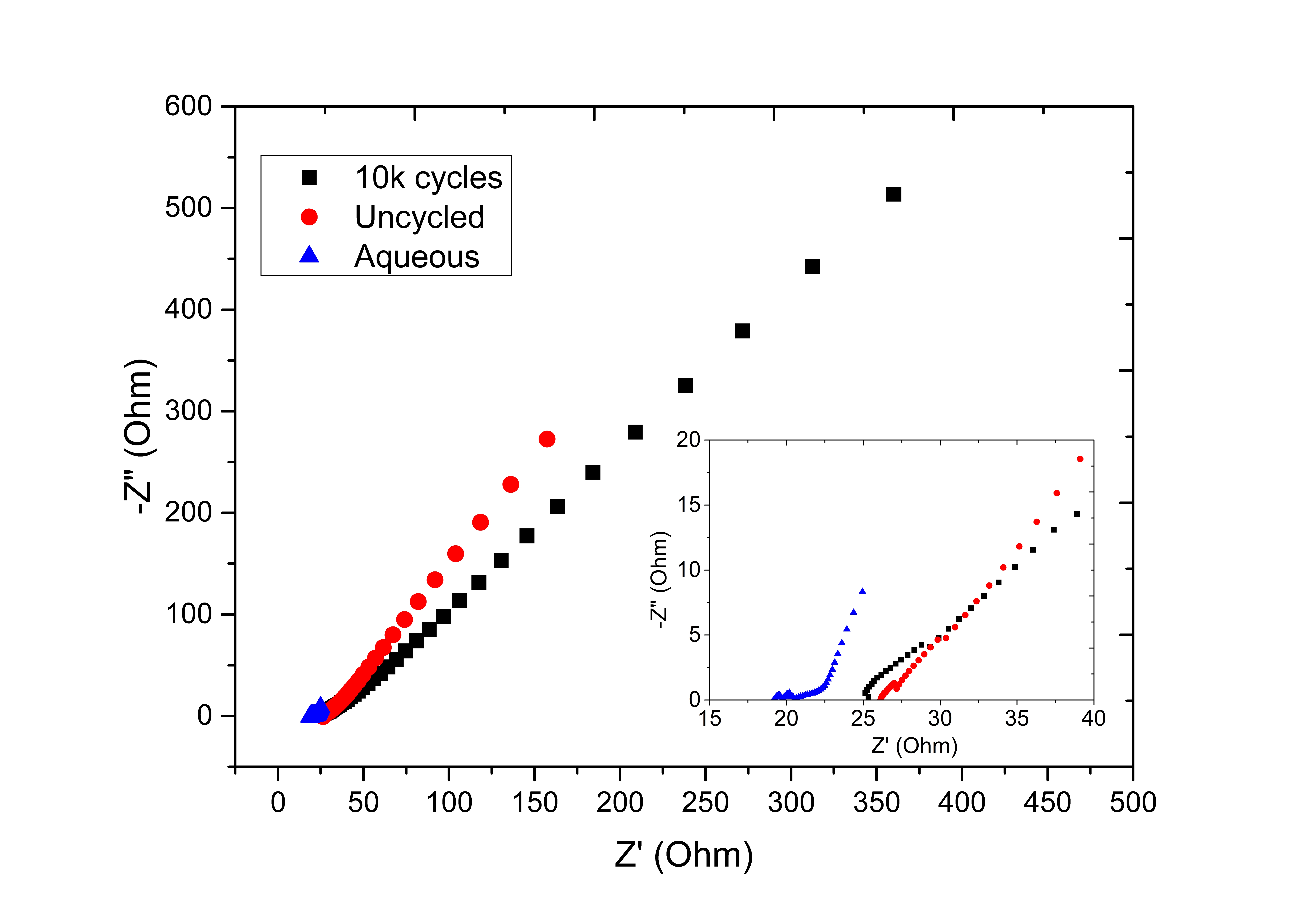}
\centering
  \caption{Electrochemical impedance spectroscopy data for the cells described. All scans were done at an applied potential of 0V with an AC amplitude of 5 mV between 0.1 and 10,000 Hz. The internal resistance of the aqueous cell is slightly lower than that of the microemulsion electrolyte due to the EEI.}
  \label{fig:Pt Windows}
\end{figure}

\begin{figure}
\includegraphics[width=\textwidth]{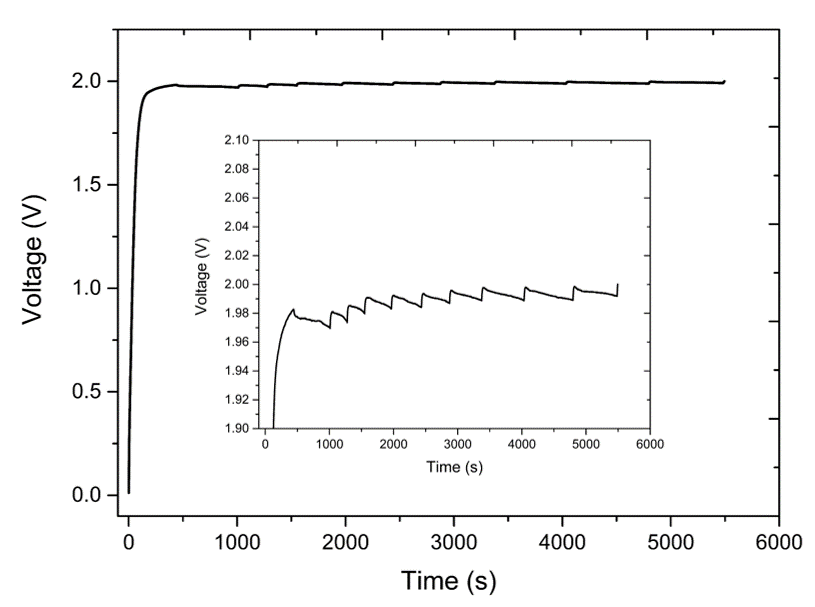}
\centering
  \caption{Galvanostatic charge curve of a cell prepared with a 0.1M aqueous NaCl electrolyte. The charging current used was 250 mA/g. Inset is a zoomed in portion of the same curve showing irregular charging behaviour indicative of solvent splitting reactions occurring inside the cell. The cell were never able to achieve 2V.}
  \label{fig:Pt Windows}
\end{figure}

\begin{figure}
\includegraphics[width=\textwidth]{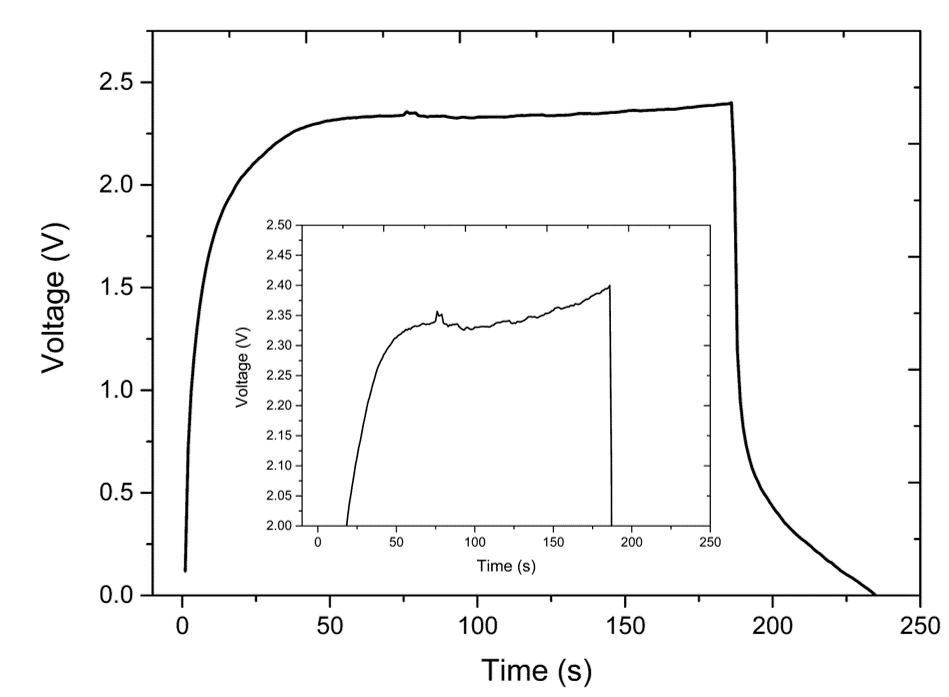}
\centering
  \caption{Galvanostatic charge discharge curve of a cell prepared with a 4.9 wt\% aqueous SDS electrolyte. The charging current used was 250 mA/g. Inset is a zoomed in portion of the same curve showing irregular charging behaviour indicative of solvent splitting reactions occurring inside the cell. The cell was able to achieve a voltage of 2.4V eventually and complete a discharge cycle, however, due to the irregularities observed in the charging curve the experiment was not continued. The IR drop was also substantial at around 500 mV which is much greater than that observed for the SDS microemulsion cells, which was around 50mV}
  \label{fig:SDS CDC}
\end{figure}

\begin{figure}
\includegraphics[width=\textwidth]{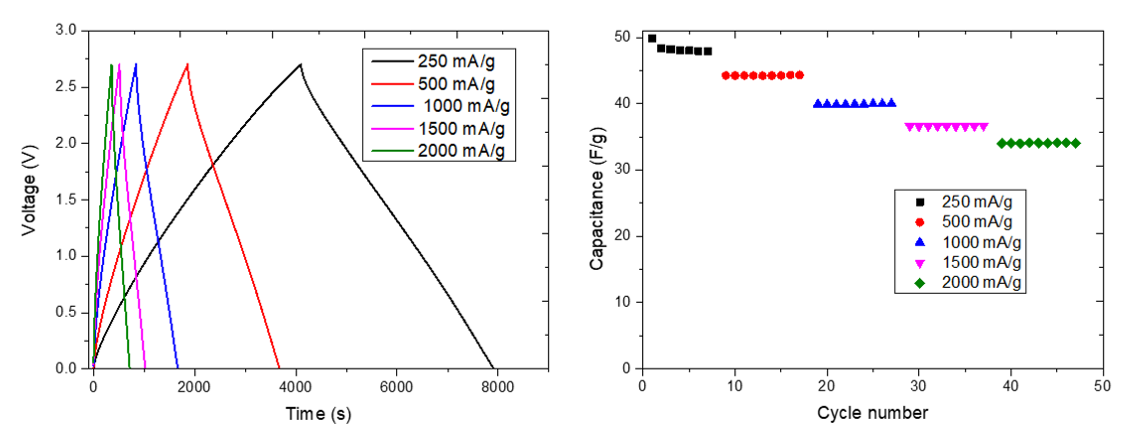}
\centering
  \caption{Results of an cell constructed using 1M TBAB in acetonitrile as an electrolyte. The rate capability of these cells were higher than that of teh ME electrolyte however the capacitance at 500 mA/g was similar. Unlike the ME cells, these cells were prepared in a nitrogen glovebox}
  \label{fig:SDS CDC}
\end{figure}

\end{document}